# GENERALIZED SCALING OF THE TURBULENCE STRUCTURE IN WALL-BOUNDED FLOWS

T.-W. Lee and J.E. Park
Department of Mechanical and Aerospace Engineering
Arizona State University
Tempe, Arizona, USA
attwl@asu.edu

**ABSTRACT**

Scaling of the Reynolds stresses has been sought by many researchers, since it provides a template of universal dynamical patterns across a range of Reynolds numbers. Various statistical and normalization schemes have been attempted, but without complete or convincing similarity properties. Our prior work on the transport processes in wall-bounded flows point toward self-similarity in the gradient space, where the first and second derivatives of the Reynolds stress components exhibit universal scaling across the entire boundary layer. This scaling is extendable to compressible flows. Finally, a universal, integral scaling for the mean velocity profiles is discovered and presented.

**Introduction**

Scaling of the mean velocity (U) profiles in turbulent flows serves a useful function of collapsing the momentum structure. Then, expansion or "stretching" by an appropriate multiplicative factor recovers the profile at other Reynolds numbers, e.g. in boundary-layer and free-jet flows. Similar patterns are sought in turbulence structures, for example the diagonal components of the Reynolds stress tensor ($u'^2$, $v'^2$, $w'^2$). In addition to replicating the profiles at arbitrary Reynolds numbers, complete scaling would lead to insights on the dynamics and origins of the turbulent flow structure,

Attempts at finding similarity in the turbulence profiles in wall-bounded flows have been mostly at the "surface" level, i.e. in the root variables (e.g. $u'^2$ as a function $y+$) and their higher statistical moments (Marusic et al., 2010, Hu et al., 2020). Measurements and DNS show progressions in the $u'^2$ profiles with increasing Reynolds number (Figure 1), wherein the peak increases in magnitude while moving closer to the wall (Graham et al., 2016; Iwamoto et al., 2002; Mansour et al., 1998; Marusic et al., 2010). Interestingly, when plotted in the inner coordinates ($y+$) the peak location stays nearly constant approximately at $y+ = 15$ (Moser et al., 1999; Marusic et al., 2010; Keirsbulck et al., 2012). Although some sectional scaling rules have been reported (Smits et al., 2021; Buschmann and Gad-el-Hak, 2007), self-similarity over the entire flow domain has been elusive.

The attached eddy hypothesis is a dynamical model of the turbulence structure in the so-called "logarithmic region". In that regard, it also suggests a method to generate a universal profile, although the transposition from the hypothesis to the actual statistics are not as straightforward (Hu et al., 2020). The model assumes that a hierarchy of eddies near the wall would lead to a logarithmic dependence for the normal components of the Reynolds stress (Marusic et al., 2010; Marusic and Monty, 2019; Hu et al., 2020), but this argument is applicable only in the descending segment of the profile (Adrian, 2010; Marusic and Monty, 2019). In the gradient ($d/dy+$) space, $u'^2 = -A\log(y+)+B$ would correspond to $du'^2/dy+ = -A/y+$, which is plotted in Figure 2 along with DNS data from Iwamoto et al. (2002) and Graham et al. (2016) for turbulent channel flows. In Figure 2, overlap of the logarithmic distribution with the $du'^2/dy+$ profiles is rather brief, while noting the fact that log scale is used for $y+$. Whether or not the attached eddy hypothesis and its dynamical prescriptions are correct in the logarithmic region, it is not of much use in scaling or explaining other prominent features in $u'^2$ (e.g. near-wall peak shapes) or its relations with other Reynolds stress components.

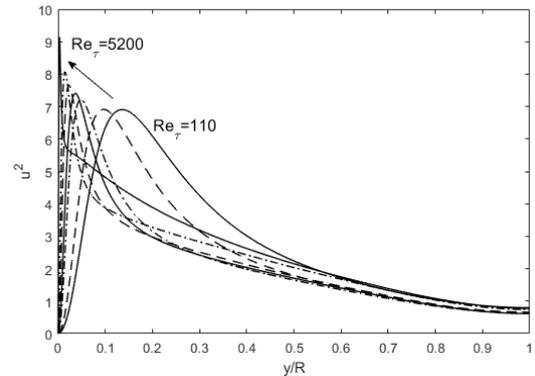

Figure. 1. Progression of $u'^2$ profiles with increasing Reynolds number in channel flows. The DNS data from Iwamoto et al. (2002) and Graham et al. (2016) are used.

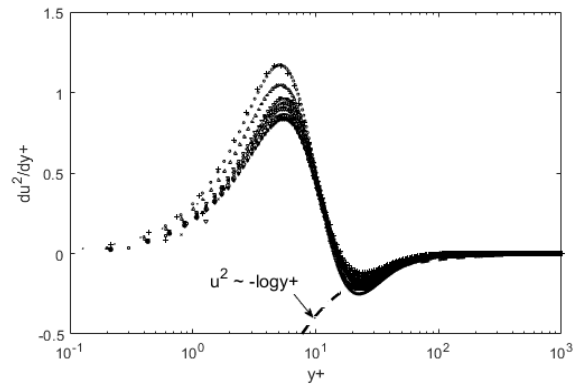

Figure. 2. A look at the gradient structure, $du'^2/dy+$. Logarithmic dependence (attached eddy model) is plotted as a dashed line. The DNS data from Iwamoto et al. (2002) and Graham et al. (2016) are used.

Figure 2 actually starts to illustrate the turbulence dynamics involving $u'^2$. When plotted as gradients ($du'^2/dy+$), the profiles now become self-similar, with a common zero-crossing point at $y+ \sim 15$ (peak location for $u'^2$). With respect to this peak location, both the near-wall positive $du'^2/dy+$ ($u'^2$ ascending) and mid-layer negative slope ($du'^2/dy+ < 0$) sections now exhibit self-similarity for the entire range of Reynolds number ($Re_\tau = 110\text{-}5200$) plotted. These scaling characteristics distinctly point to the advantages of examining turbulence structure in the gradient ($d/dy+$) space. In this work, we show that scaling can be found at the first and second gradient ($d/dy+$, $d^2/dy+^2$) levels for the Reynolds stress components. Implications on the origin of such turbulence structure in wall-bounded flows are also discussed, along with transport relationships between $u'^2$, $v'^2$, and $u'v'$.

**Scaling of the Reynolds Stresses**

If we take the first gradient of $u'^2$ profiles, then self-similarity is found across a large range of Reynolds numbers and different wall-bounded (flat-plate and channel) flows, as was shown in Figure 2. The fixed peak location ($y+ \sim 15$) serves as a pivot (zero-crossing) point for positive and negative segments. Also, the maxima and minima in the gradients vary with the Reynolds number (Lee, 2021a; Lee, 2021b), asymmetrically (steeper for the positive segments) but both in a monotonic manner so that scaling factors can be introduced to collapse the profiles. The maximum $u'^2$ variation with the Reynolds number has been correlated using DNS data (Keirsbulck et al., 2012). Similarly, maxima (peak) and minima (nadir) for $du'^2/dy+$ can be tabulated as a function of the Reynolds number. We have used the DNS data of Iwamoto et al. (2002) and Graham et al. (2016) for channel flows ($Re_\tau = 110\text{-}5200$), and also of Spalart (1998) for boundary-layer flow over a flat plate ($Re_\tau = 300 – 1410$), for the turbulence profiles including evaluations of the maxima and minima in $du'^2/dy+$ (Figure 2). The profiles in Figure 3 are thus scaled in the $y+$ axis through the $d/dy+$ operation, and normalized by the absolute values of the extrema (maxima/minima), $(du'^2/dy+)_{ext}$, in the vertical direction. Upon doing so, we can see that in Figure 3 the collapse of the profiles is nearly perfect, universal across the Reynolds number for both the channel and boundary-layer flows, and covers the entire flow width. Segmented, partial scaling rules are no longer necessary with the universal gradient ($d/dy+$) scaling. We should be also able to deduce the turbulence dynamics from this self-similarity characteristic, as discussed in the next section.

Similar patterns are observed for $v'^2$ and $u'v'$, except at the *second-gradient* level (Figures 4 and 5). As shown in Figures 3-5, the gradient profiles all collapse when properly normalized by the peak/nadir heights of the second gradients. Thus, a profile at a reference Reynolds number can be used to reconstruct $u'^2$, $v'^2$, or $u'v'$ at any other $Re_\tau$, by appropriate "stretching" (Lee, 2021a; Lee, 2021b).

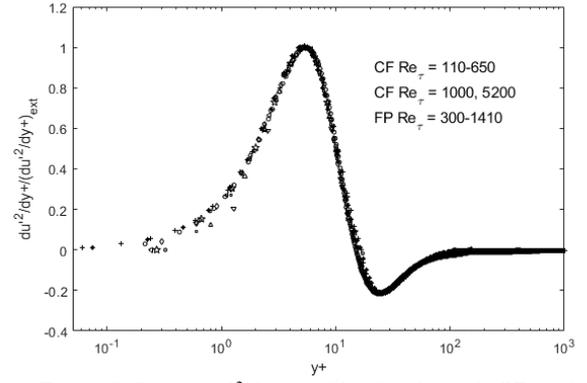

Figure 3. Scaled $du'^2/dy+$ profiles for channel (CF) and boundary-layer (FP) flows. The DNS data from Graham et al. (2016) and Spalart (1998) are used.

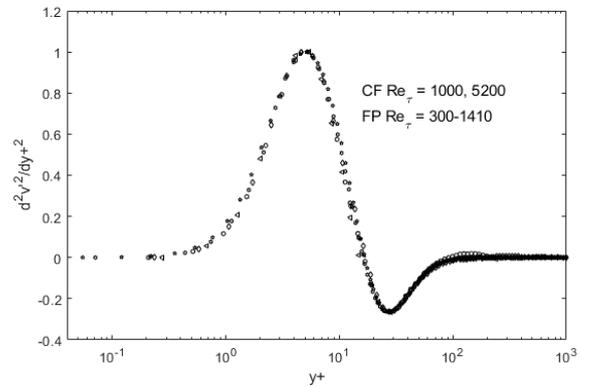

Figure. 4. . Scaled $d^2v'^2/dy+^2$ profiles for channel (CF) and boundary-layer (FP) flows. The DNS data from Graham et al. (2016) and Spalart (1998) are used. The data have been normalized by the absolute values of the maxima and minima, similar to Figure 3.

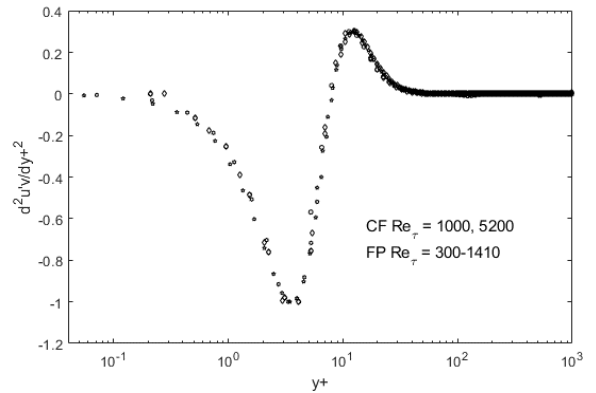

Figure 5. Scaled $d^2u'v'/dy+^2$ profiles for channel (CF) and boundary-layer (FP) flows. The DNS data from Graham et al. (2016) and Spalart (1998) are used. The data have been normalized by the absolute values of the maxima and minima, similar to Figure 3.

The results in the above figures (Figs. 3-5) show that the gradient scaling is complete over the flow width and universal for wall-bounded flows for a large range of Reynolds numbers. Recent examinations of the turbulence profiles in adverse pressure-gradient boundary layers indicate that the self-similarity also exists in other turbulence configurations (Lee and Park, 2023).

**Compressible Flows**

The self-similarity exhibited in the previous section can be extended to compressible turbulent flows using DNS data by Gerolymos and Vallet (2023), as shown in Figs. 6-8. Fig. 6 is the scaled d<$\rho u''^2$>/dy+ profiles for the near- and far-wall structures in Fig. 6a and b, respectively. Due to the expansion effect in compressible flows, density-weighted averaging and y-coordinate (y*) is used (Gerolymos and Vallet, 2023). Then, similar collapse of the profiles occurs in the first-gradient space, albeit differently for near- and far-wall regions. For u''v'' profiles, similarity is again found for the first gradients, in Fig. 7, while $v''^2$ follows the second-gradient scaling. $d^2$<$\rho v''^2$>/dy+$^2$ structure is also somewhat peculiar in that in the far-wall region the profiles deviate from the similarity rule (second gradients collapse).

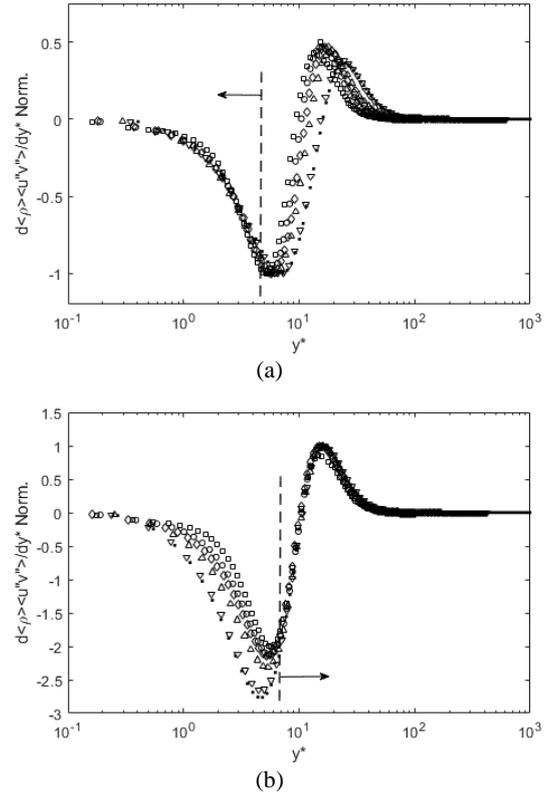

(a)

(b)

Figure 7. Scaled d<$\rho u''v''$>/dy+ profiles for the near-wall (a) and far regions (b). The DNS data from Gerolymos and Vallet (2023) is used, for compressible plane channel flows.

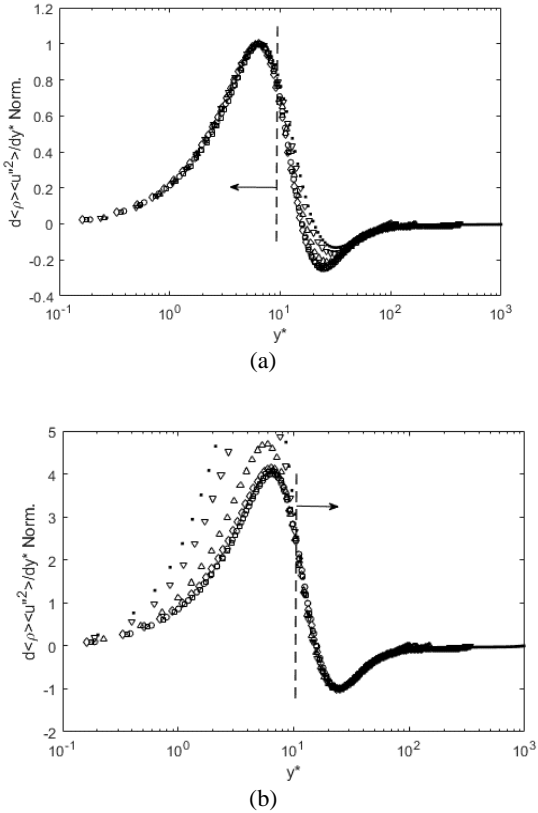

(a)

(b)

Figure 6. Scaled d<$\rho u''^2$>/dy+ profiles for the near-wall (a) and far regions (b). The DNS data from Gerolymos and Vallet (2023) is used, for compressible plane channel flows.

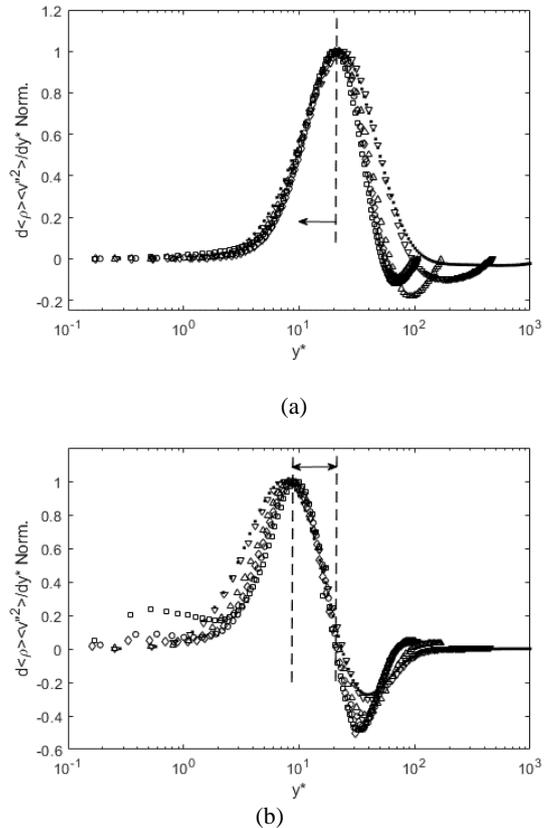

(a)

(b)

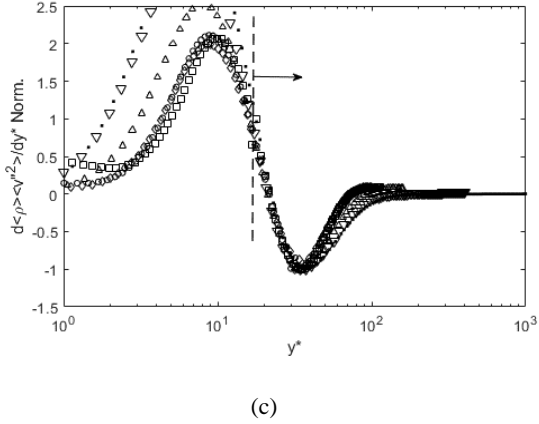

(c)

Figure 8. Scaled $d^2\langle\rho v''^2\rangle/dy+^2$ structure for the near-wall (a), intermediate (b) and far regions (c). The DNS data from Gerolymos and Vallet (2023) is used, for compressible plane channel flows.

**Mean Velocity Scaling**

Mean velocity scaling has been subject to intense debate. Here, we present a universal, integral scaling that collapses the velocity profiles across the entire flow width (wall to centerline). For example, the U+ profiles are plotted in Fig. 9 for the compressible, plane channel flows using DNS data (Gerolymos and Vallet, 2023). Except near the wall and in the narrow logarithmic region, the mean velocity profiles are dispersed. There may be some scaling rules to perhaps narrow the dispersion. However, a simple integral operation collapses the mean velocity profiles, as shown in Fig. 10.

$$\text{Integrated U}^+ = \int_0^{y^+} U(y^+)dy^+ \quad (1)$$

Similar to the gradient scaling above, this integral scaling for the mean velocities appears to be general, as applied to incompressible flows in Fig. 11. There are two segments in the integral space, so that if we parameterize them we will have general mean velocity scaling for the entire flow width in turbulent flows.

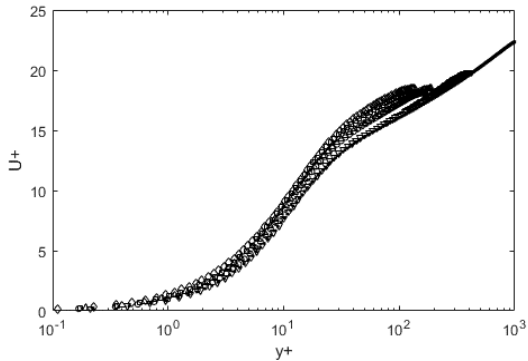

Figure 9. Mean velocity profile in compressible plane channel flows. The DNS data from Gerolymos and Vallet (2023) is used.

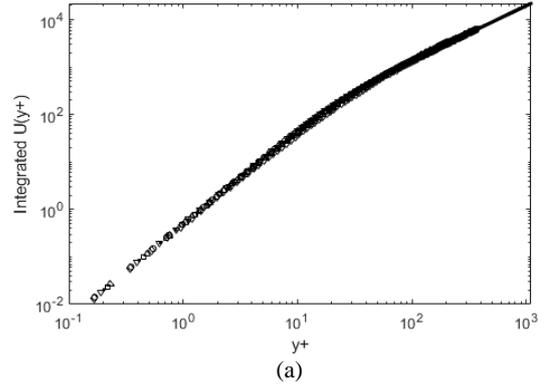

(a)

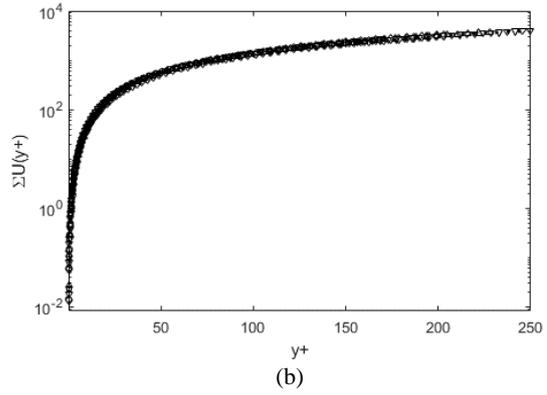

(b)

Figure 10. Integrated U+ in logarithmic (a) and linear (b) y+ coordinates. The DNS data from Gerolymos and Vallet (2023) is used.

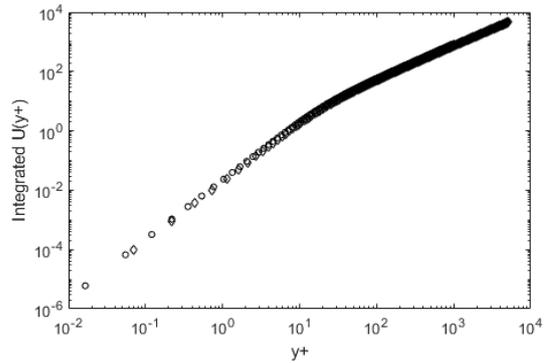

Figure 11. Integrated U+ scaling for incompressible channel flows. The DNS data from Graham et al. (2016) is used.

**Concluding Remarks**

The dissipation scaling in the turbulence structure is discussed, along with its dynamical origins. In the gradient space (d/dy+, $d^2/dy+^2$), the self-similarity is complete and universal, in incompressible and compressible wall-bounded flows. In addition, a universal, integral scaling for the mean velocity structure is also discovered.

**REFERENCES**

Adrian, R.J., 2010, Closing in on models of wall turbulence, Science **329**, 155–156.

Graham, J., Kanov, K, Yang, X.I.A., Lee, M.K., Malaya, N, Lalescu, C.C., Burns, R., Eyink, G, Szalay, A, Moser, R.D., and Meneveau, C., 2016, A Web Services-accessible database of turbulent channel flow and its use for testing a new integral wall model for LES, Journal of Turbulence, 17(2), 181-215.

Gerolymos, G.A. and Vallet, I., 2023, Scaling of pressure fluctuations in compressible plane channel flows, Journal of Fluid Mechanics, Vol. 958, A19.

..Hu, B., Xiang, I.A.Y. and Zheng X., 2020, Wall-attached eddies in wall-bounded turbulent flows, Journal of Fluid Mechanics, Vol. 194, pp. 15-44.

Iwamoto, K., Sasaki, Y., Nobuhide K., 2002, Reynolds number effects on wall turbulence: toward effective feedback control, International Journal of Heat and Fluid Flows, 23, 678-689.

Keirsbulck, L., Fourrie, G., Labraga, L., and Gad-el-Hak, M., 2012, Scaling of statistics in wall-bounded turbulent lows, Comptes Rendus Mecanique, 340, 420-433.

Lee, T.-W., 2020a, Lagrangian Transport Equations and an Iterative Solution Method for Turbulent Jet Flows, Physica D, 132333.

Lee, T.-W., A Generalizable Theory of the Reynolds Stress, 2020b, arXiv:2006.01634.

Lee, T.-W., 2021a, Dissipation scaling and structural order in turbulent channel flows, Physics of Fluids, 33, 5, 055105.

Lee, T.-W., 2021b, Asymmetrical Order in Wall-Bounded Turbulent Flows, Fluids, 6(9), 329.

Lee, T.-W., 2023, Dynamics and scaling of the Reynolds stress in adverse pressure-gradient turbulent boundary layer flows, arXiv:2310.19723, also to be submitted to a journal.

Mansour, N.N., Kim, J. and Moin, P., 1998, Reynolds-stress and dissipation-rate budgets in a turbulent channel flow, Journal of Fluid Mechanics, Vol. 194, pp. 15-44.

Marusic, I., McKeon, B.J., Monkewitz, P.A., Nagib, H.M., Smits, A.J., and Sreenivasan, K.R., 2010, Wall-bounded turbulent flows at high Reynolds numbers: Recent advances and key issues, Physics of Fluids, 22, 065103.

Marusic, I. and Monty, J.P., 2019, Attached eddy model of wall turbulence, Annual Review of Fluid Mechanics, 51:49-74.

Moser, Robert D., John Kim, and Nagi N. Mansour, 1999, Direct numerical simulation of turbulent channel flow up to $Re_t$= 590, Physics of Fluids, 1999, 11,4, pp. 943-945.

Smits, A.J., Hultmark, M., Lee, M., Pirozzoli, S., and Wu, X., 2021, Reynolds stress scaling in the near-wall region of wall-bounded flows, Journal of Fluid Mechanics, 926, A31.

Spalart, P.R., 1988, Direct simulation of a turbulent boundary layer up to Re=1410, Journal of Fluid Mechanics, Vol. 187, pp. 61-77.

Tennekes, H. and Lumley, J.L., A first course in turbulence, MIT press, 1972.